\documentclass{jfm}

\usepackage{graphicx}

\newcommand{\be}{\begin{equation}}
\newcommand{\ee}{\end{equation}}
\newcommand{\bea}{\begin{eqnarray}}
\newcommand{\eea}{\end{eqnarray}}
\newcommand{\bc}{\begin{center}}
\newcommand{\ec}{\end{center}}
\newcommand{\btab}{\begin{tabular}}
\newcommand{\etab}{\end{tabular}}

\let\oldepsilon\epsilon
\let\epsilon\varepsilon
\let\varepsilon\oldepsilon
\let\oldphi\phi
\let\phi\varphi
\let\varphi\oldphi

\title{Relaxation of a dewetting contact line\\ Part 1: A full-scale hydrodynamic calculation}

\author[ J.H. Snoeijer, B. Andreotti, G. Delon and M. Fermigier] 
{J\ls A\ls C\ls C\ls O\ns H.\ns S\ls N\ls O\ls E\ls I\ls J\ls E\ls R, 
B\ls R\ls U\ls N\ls O\ns A\ls N\ls D\ls R\ls E\ls O\ls T\ls T\ls I, 
G\ls I\ls L\ls E\ls S\ns D\ls E\ls L\ls O\ls N \and M\ls A\ls R\ls C\ns F\ls E\ls R\ls M\ls I\ls G\ls I\ls E\ls R}

\affiliation{Physique et M\'ecanique des Milieux 
H\'et\'erog\`enes, ESPCI, 10 rue Vauquelin, 75231 Paris  
Cedex 05, France}

\date{\today}
\pubyear{2000}
\volume{???}
\pagerange{??--??}
\date{\today}
\begin{document}
\maketitle

\begin{abstract}
{The relaxation of a dewetting contact line is investigated theoretically in the so-called "Landau-Levich" geometry in which a vertical solid plate is withdrawn from a bath of partially wetting liquid. The study is performed in the framework of lubrication theory, in which the hydrodynamics is resolved at all length scales (from molecular to macroscopic). We investigate the bifurcation diagram for unperturbed contact lines, which turns out to be more complex than expected from simplified 'quasi-static' theories based upon an apparent contact angle. Linear stability analysis reveals that below the critical capillary number of entrainment, ${\rm Ca}_c$, the contact line is linearly stable at all wavenumbers. Away from the critical point the dispersion relation has an asymptotic behaviour $\sigma\propto |q|$ and compares well to a quasi-static approach. Approaching ${\rm Ca}_c$, however, a different mechanism takes over and the dispersion evolves from $\sim |q|$ to the more common $\sim q^2$. These findings imply that contact lines can not be treated as universal objects governed by some effective law for the macroscopic contact angle, but viscous effects have to be treated explicitly.}
\end{abstract}

\section{Introduction}

Wetting and dewetting phenomena are encountered in a variety of environmental and technological contexts, ranging from the treatment of plants to oil-recovery and coating. Yet, their dynamics can not be captured within the framework of classical hydrodynamics --~with the usual no-slip boundary condition on the substrate~-- since the viscous stress diverges at the contact line (\cite{HS71,DDD74}). The description of moving contact lines has remained a great challenge, especially because it involves a wide range of length scales. In between molecular and millimetric scales, the strong viscous stresses are balanced by capillary forces. In this zone, the slope of the free surface varies logarithmically with the distance to the contact line so that the interface is strongly curved, even down to small scales (\cite{V76,C86}). Ultimately, the intermolecular forces due to the substrate introduce the physical mechanism that cuts off this singular tendency (\cite{V76,C86,DG86,B95,PP00}). 

A popular theoretical approach has been to assume that all viscous dissipation is localized at the contact line, so that macroscopically the problem reduces to that of a static interface that minimizes the free energy. In such a quasi-static approximation one does not have to deal explicitly with the contact line singularity: the dynamics is entirely governed by an {\em apparent contact angle} $\theta_a$ that serves as a boundary condition for the interface (\cite{V76,C86,JdG84,GR01b,NV03}). This angle is a function of the capillary number ${\rm Ca}=\eta U/\gamma$, which compares the contact line velocity $U$ to the capillary velocity $\gamma/\eta$, where $\gamma$ and $\eta$ denote surface tension and viscosity. Since the dissipative stresses are assumed to be localized at the contact line, viscous effects will modify the force balance determining the contact angle. This induces a shift with respect to the equilibrium value $\theta_e$. Within this approximation, the difficultly of the contact line problem is  hidden in the relation $\theta_a({\rm Ca})$, which depends on the mechanism releasing the singularity. While it is agreed upon that the angle increases with ${\rm Ca}$ for advancing contact lines and decreases in the receding case, there are many different theories for the explicit form (\cite{V76,C86,DG86,B95}). 

\begin{figure}
\includegraphics{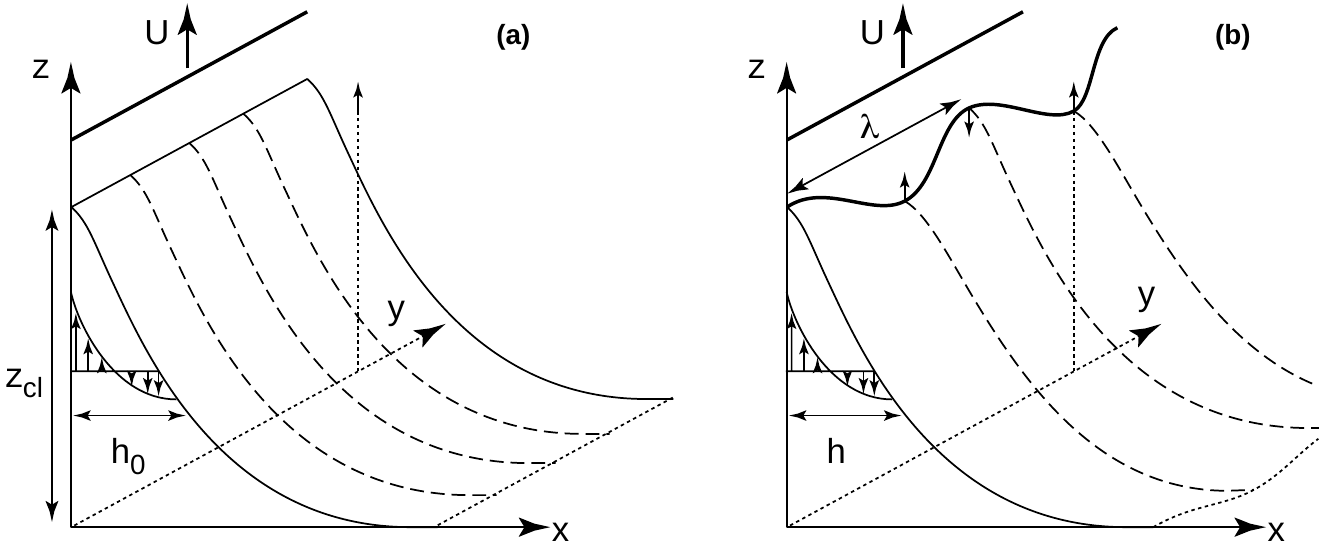} 
\centering 
\caption{(a) A standard geometry to study contact line dynamics is that of a vertical solid plate withdrawn from a bath of liquid with a constant velocity $U$. The position of the contact line is indicated by $z_{\rm cl}$. (b) In this paper we study the relaxation of transverse perturbations of contact lines, by computing the evolution of the interface profile $h(z,y,t)$.} 
\label{fig.setup} 
\end{figure} 

Experimentally, however, it has turned out to be very difficult to discriminate between the various theoretical proposals (\cite{H75,LG05,RDAL05}). All models predict a nearly linear scaling of the contact angle in a large range of ${\rm Ca}$ and the prefactor is effectively an adjustable parameter (namely the logarithm of the ratio between a molecular and macroscopic length). Differences become more pronounced close to the so-called forced wetting transition: it is well known that the motion of {\em receding} contact lines is limited by a maximum speed beyond which liquid deposition occurs~(\cite{BR79,DG86}). An example of this effect is provided by drops sliding down a window. At high velocities, these develop singular cusp-like tails that can emit little droplets (\cite{PFL01,LG05}). Similarly, solid objects can be coated by a non-wetting liquid when withdrawn fast enough from a liquid bath (\cite{BR79,Q91,S91}), see Fig.~\ref{fig.setup}a. Above the transition, a capillary ridge develops (\cite{SDFA06}) that eventually leaves a Landau-Levich film of uniform thickness (\cite{LL42}). 

An important question is to what extent a quasi-static approximation, in which dissipative effects are taken localized at the contact line, are able to describe these phenomena. Only recently, the problem has been addressed by using a fully hydrodynamic model that properly incorporates viscous effects at all length scales (\cite{H01,E04,E05}). It was found that stationary meniscus solutions cease to exist above a critical value ${\rm Ca}_c$, due to a matching problem at both ends of the scale range: the highly curved contact line zone and the macroscopic flow (\cite{E04,E05}). An interesting result of this work is that both the value of ${\rm Ca}_c$ and the emerging $\theta_a({\rm Ca})$ are not universal: these depend on the inclination at which the plate is withdrawn from the liquid reservoir. Hence, the large scale geometry of the interface {\em does} play a role and the dynamics of contact lines can not be captured by a single universal law for $\theta_a({\rm Ca})$. 

\begin{figure}
\includegraphics{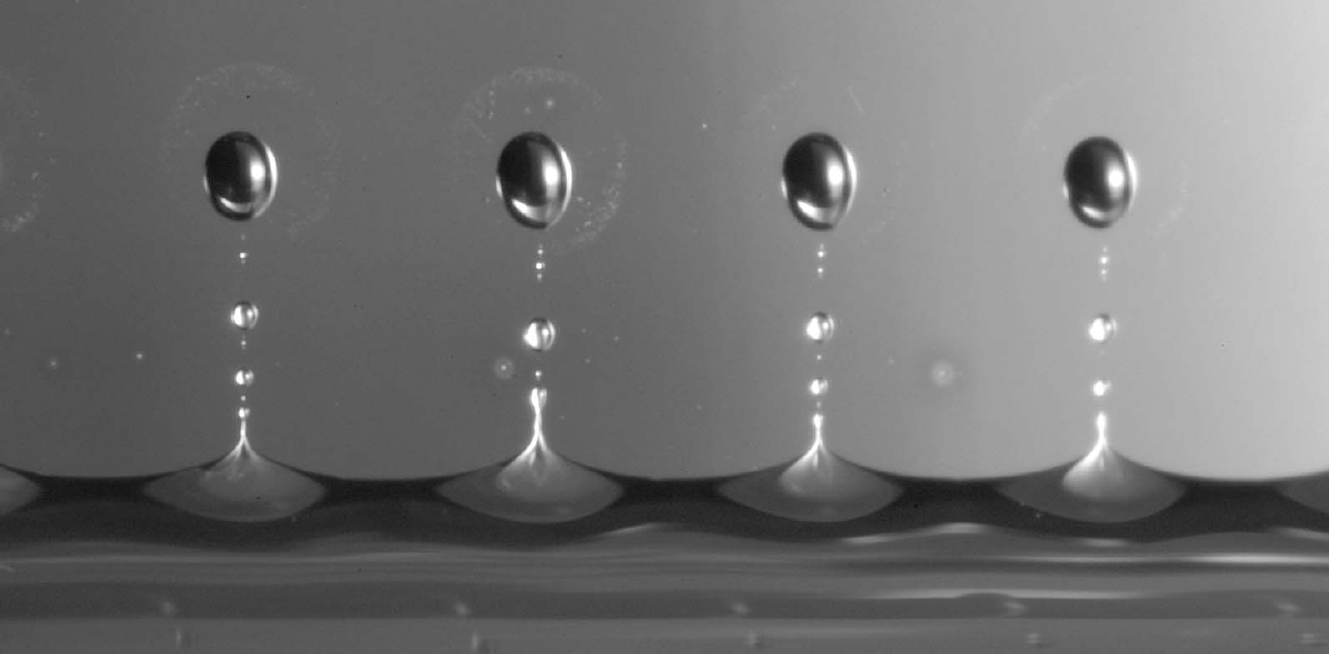} 
\centering 
\caption{Experimental realization of contact line perturbations. The contact line is deformed by "wetting defects" on the partially wetting plate. The narrow connection between the defect and the bath undergoes a Rayleigh-Plateau-like instability, leaving a periodically deformed contact line.} 
\label{fig.manip} 
\end{figure}

Golestanian and Raphael identified another sensitive test to discriminate contact line models (\cite{GR01a,GR01b}). They considered the relaxation of dewetting contact lines perturbed at a well-defined wave number $q$, as shown in Fig.~\ref{fig.setup}b. This can be achieved experimentally by introducing wetting defects on the solid plate, separated by a wavelength $\lambda$ (\cite{OV91b}). As can be seen in Fig.~\ref{fig.manip}, these defects create a nonlinear perturbation when passing through the contact line, but eventually the relaxation occurs along the Fourier mode with $q=2\pi/\lambda$ (\cite{JFM2}). Using a quasi-static theory, Golestanian and Raphael predict that the perturbations decay exponentially $\sim e^{-\sigma t}$ with a relaxation rate
\begin{equation}\label{woef}
\sigma = |q| \frac{\gamma}{\eta}\, f({\rm Ca})~,
\end{equation}
where $f({\rm Ca})$ is very sensitive to the form of $\theta_a({\rm Ca})$. Their theory is built upon the work by \cite{JdG84}, who already identified the scaling proportional to $|q|$ for static contact lines (${\rm Ca}=0$). Ondar\c{c}uhu and Veyssi\'e experimentally confirmed this $|q|$ dependence in the limit of ${\rm Ca}=0$ (\cite{OV91b}), while more recently, it has been argued that this scaling should saturate to the inverse capillary length $l_\gamma=\sqrt{\gamma/\rho g}$ in the large wavelength limit (\cite{NV03}). However, an intriguing and untested prediction for the {\em dynamic} problem is that the relaxation times diverge when approaching forced wetting transition at ${\rm Ca}_c$. This "critical" behavior should occur at all length scales and is encountered in the prefactor $f({\rm Ca})$, which vanishes as ${\rm Ca} \rightarrow {\rm Ca}_c$.

In this paper we perform a fully hydrodynamic analysis of perturbed menisci when a vertical plate is withdrawn from a bath of liquid with a velocity $U$ (Fig.~\ref{fig.setup}). Using the lubrication approximation it is possible to take into account the viscous dissipation at all length scales, from molecular (i.e. the slip length) to macroscopic. We thus drop the assumptions of quasi-static theories and describe the full hydrodynamics of the problem. The first step is to compute the unperturbed meniscus profiles as a function of the plate velocity. We show that these basic solutions undergo a remarkable series of bifurcations that link the effect that stationary cease to exist beyond ${\rm Ca}_c$(\cite{E04}), to the recently observed upward propagating fronts beyond the transition (\cite{SDFA06}). Then we study the dispersion of contact line perturbations through a linear stability analysis. Our main findings are: (i) the relaxation time for the mode $q=0$ scales as $|{\rm Ca}-{\rm Ca}_c|^{-1/2}$; (ii) finite wavelength perturbations always decay in a finite time even right at the critical point; (iii) the scaling $\sigma \propto |q|$ proposed by Eq.~(\ref{woef}) breaks down when approaching ${\rm Ca}_c$. These results illustrate the limitations of simplified theories based upon an apparent contact angle. 

The paper is organized as follows. In Sec.~\ref{sec.quasistatic} we summarize the results from a quasi-static theory and generalize the work by Golestanian \& Raphael. The heart of the paper starts in Sec.~\ref{sec.hydro} where we formulate the hydrodynamic approach and compute the bifurcation diagram of the base solutions. After addressing technical points of the linear stability analysis in Sec.~\ref{sec.linstab}, we present our numerical results for the dispersion relation in Sec.~\ref{sec.results}. The paper closes with a discussion in Sec.~\ref{sec.discussion}.

\section{Results from quasi-static theory}\label{sec.quasistatic}

We briefly revisit the quasi-static approach to contact line perturbations, which will serve as a benchmark for the full hydrodynamic calculation starting in Sec.~\ref{sec.hydro}. The results below are based upon the analysis of Golestanian \& Raphael, which has been extended to long wavelengths and large contact angles. 

\subsection{Short wavelengths: $q l_\gamma \gg 1$}\label{sec.staticshort}

\begin{figure} 
\includegraphics{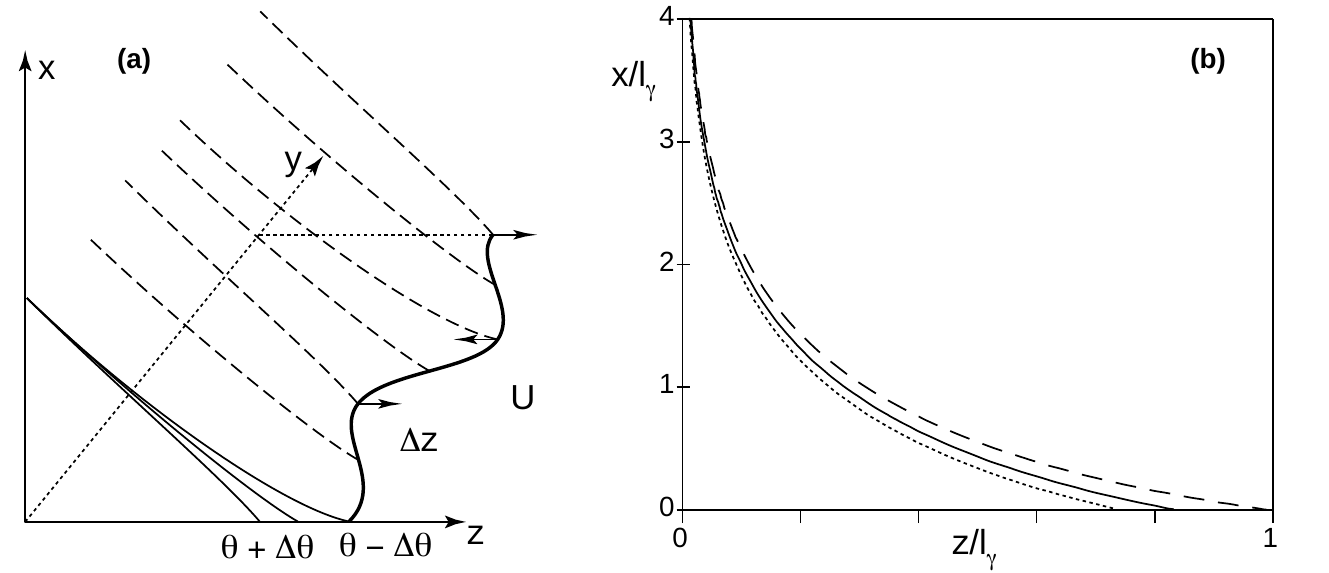} 
\centering 
\caption{(a) Macroscopic representation of the interface shape near a perturbed contact line. The advanced part of the contact line has a smaller apparent contact angle than the unperturbed $\theta_a$, and will thus have a higher speed with respect to the plate. This will decrease the amplitude of the perturbation. (b) Cross-sections of the perturbed interface profile along $z$. Note that the interface joins the static bath at $z=0$.} 
\label{fig.sketch} 
\end{figure}

At distances well below the capillary length, $l_\gamma=\sqrt{\gamma/\rho g}$, we can treat the unperturbed profiles as a straight wedge of angle $\theta_a$. Perturbations should not affect the total Laplace pressure, and hence not the total curvature of the free interface. The interface will thus be deformed as sketched in Fig.~\ref{fig.sketch}a: the advanced part of the contact line has a smaller apparent contact angle than the unperturbed $\theta_a$. According to $\theta_a(\rm Ca)$, such a smaller angle corresponds to a higher velocity with respect to the plate, hence the perturbation will decay. From this argument one readily understands that the rate of relaxation $\sigma$, depends on how a variation of $\theta$ induces a variation of ${\rm Ca}$, and thus involves the derivative $d{\rm Ca}/d \theta_a$ (\cite{GR03}). 

Working out the mathematics, see Appendix~\ref{app.quasistatic}, we find

\begin{equation}
\label{qfinitestatic}
\frac{\eta \sigma}{|q|\gamma} = - \frac{\tan \theta_a}{\cos \theta_a} \left( \frac{d \tan \theta_a}{d{\rm Ca}} \right)^{-1}. 
\end{equation}
This implies that the timescale for the relaxation is set by the length $q^{-1}$ and the capillary velocity $\gamma/\eta$, where $\gamma$ represents surface tension and $\eta$ is the viscosity. We therefore introduce

\begin{equation}\label{sigmainf}
\sigma_\infty({\rm Ca}) = \lim_{ql_\gamma \rightarrow \infty} \frac{\sigma}{q l_\gamma} \, \frac{\eta l_\gamma}{\gamma}~,
\end{equation}
which will be used later on to compare to the hydrodynamic calculation in the limit of large $q$. 

\subsection{Large wavelengths: $ql_\gamma \ll 1$}

When considering modulations of the contact line with $1/q$ of the order of the capillary length, one can no longer treat the basic profile as a simple wedge. Instead, one has to invoke the full profile $h_0(z)$ and the results for $\sigma$ are no longer geometry independent (see also \cite{SOK87}). For the geometry of a vertical plate immersed in a bath of liquid, we can characterize the profiles by the "meniscus rise", indicating the position of the contact line $z_{\rm cl}$ above the liquid bath (Fig.~\ref{fig.setup}). This is directly related to the contact angle as (\cite{LL59})
\begin{equation}
\label{zcl}
z_{\rm cl} = \pm\,l_\gamma \sqrt{2(1-\sin \theta_a)},
\end{equation}
where the sign depends on whether $\theta_a<\pi/2$ (positive), $\theta_a> \pi/2$ (negative). In fact, this relation is often used to experimentally determine $\theta_a({\rm Ca})$, since the meniscus rise $z_{\rm cl}({\rm Ca})$ can be measured more easily than the slope of the interface.

We now consider the relaxation rate $\sigma_0$ for perturbations with $q=0$. Such a perturbation corresponds to a uniform translation of the contact line with $\Delta z$. Using the empirical relation $z_{\rm cl}({\rm Ca})$, we can directly write
\begin{equation}
\frac{d\Delta z}{dt} =  - \frac{\gamma}{\eta}\, \Delta {\rm Ca} = -\frac{\gamma}{\eta}\, \left( \frac{d z_{\rm cl}}{d{\rm Ca}} \right)^{-1} \Delta z.
\end{equation}
Hence,
\begin{equation}\label{q0fromzcl}
\frac{\eta l_\gamma \sigma_0}{\gamma} = l_\gamma \left( \frac{d z_{\rm cl}}{d{\rm Ca}} \right)^{-1}. 
\end{equation}
In terms of the contact angle, using Eq.~(\ref{zcl}), this becomes
\begin{equation}
\label{q0static}
\frac{\eta l_\gamma \sigma_0}{\gamma} = - \frac{\sqrt{2(1-\sin \theta_a)}}{|\cos^3 \theta_a|} \,\left(\frac{d \tan \theta_a}{d{\rm Ca}} \right)^{-1} .
\end{equation}
Besides some geometric factors, this result has the same structure as the relaxation for small wavelengths, Eq.~(\ref{qfinitestatic}). The crucial difference, however, is that the length scale of the problem is now $l_\gamma$ instead of $q$. Comparing the relaxation of finite wavelenghts, $\sigma_q$, with the zero mode relaxation, $\sigma_0$, we thus find
\begin{equation}
\label{ratiostatic}
\frac{\sigma_q}{\sigma_0} \simeq ql_\gamma \, g(\theta_a) \quad {\rm for} \quad  ql_\gamma \gg 1,
\end{equation}
where the prefactor $g(\theta_a)$ reads
\begin{equation}
g(\theta_a)=\frac{|\cos \theta_a | \sin \theta_a}{\sqrt{2(1-\sin \theta_a)}}.
\end{equation}
Using the definition Eq.~(\ref{sigmainf}) we thus find the quasi-static prediction

\begin{equation}\label{ratiosigmainf}
\sigma_\infty = \sigma_0 \, g(\theta_a)~.
\end{equation}

\subsection{Physical implications and predictions}\label{sec.implications}

The predictions of the quasi-static approach can be summarized by Eqs.~(\ref{qfinitestatic}), (\ref{q0fromzcl}), and~(\ref{ratiosigmainf}). The relation $\sigma \propto |q|$ was already found by \cite{JdG84}, who referred to this as the "anomalous elasticity" of contact lines. The linear dependence on $q$ contrasts with the more generic scaling $q^2$ for diffusive systems, and has been confirmed experimentally by \cite{OV91b} in the static limit,  ${\rm Ca}=0$. An interesting consequence is that the Green's function corresponding to this dispersion relation is a Lorentzian $\propto (1+[y/w(t)]^2)^{-1}$, whose width $w(t)$ grows linearly in time. The prediction is thus that a localized deformation of the contact line, similar to Fig.~\ref{fig.manip} but now for a single defect, will display a broad power-law decay along $y$. In the hydrodynamic calculation below we will identify a breakdown of this phenomenology in the vicinity of the critical point. 

On the level of the speed-angle law $\theta_a({\rm Ca})$, the wetting transition manifests itself through a maximum possible value of ${\rm Ca}$, i.e. $d\theta/d{\rm Ca}=\infty$. According to  Eqs.~(\ref{qfinitestatic}) and~(\ref{q0static}), this suggests a diverging relaxation time $\sigma^{-1}$ at all length scales. Assume that the scaling close to the maximum is $\theta_a - \theta_c \propto({\rm Ca}_c-{\rm Ca})^\beta$, where generically one would expect $\beta=1/2$. If the critical point occurs at zero contact angle, $\theta_c=0$, Eq.~(\ref{qfinitestatic}) yields a scaling $\sigma_q \propto ({\rm Ca}_c-{\rm Ca})$ for the case of large $q$. For $q=0$ or when $\theta_c\neq 0$, one finds $\sigma_q \propto ({\rm Ca}_c-{\rm Ca})^{1-\beta}$. Below we show that the mode $q=0$ indeed displays the latter scaling with $\beta=1/2$. However, the relaxation times of finite $q$ perturbations always remain finite according to the full hydrodynamic calculation, even at the critical point.

\section{Hydrodynamic theory: the basic profile $h_0(z)$}\label{sec.hydro}

This section describes the hydrodynamic theory that is used to study the relaxation problem. After presenting the governing equations, we reveal the nontrivial bifurcation diagram of the stationary solutions, $h_0(z)$, for different values of the capillary number. This allows an explicit connection between the work on the existence of stationary menisci (\cite{H01,E04,E05}), and recently observed transient states in the deposition of the Landau-Levich film (\cite{SDFA06}). All results presented below have been obtained through numerical resolution of the hydrodynamic equations using a Runge-Kutta integration method. 

\subsection{The lubrication approximation}\label{subsec.lubri}

We consider the coordinate system $(z,y)$ attached to the solid plate, as indicated in Fig.~\ref{fig.setup}. The position of the liquid/vapor interface is denoted by the distance from the plate $h(z,y,t)$. To really cover the range of length scales from molecular to millimetric, the standard approach is to describe the hydrodynamics using the lubrication approximation (\cite{ODB97}). This is a long wavelength expansion of the Stokes flow based upon  ${\rm Ca} \ll 1$, which reduces the free boundary problem to a single partial differential equation for $h(z,y,t)$. Of course, one still has to deal explicitly with the fact that viscous forces tend to diverge as $h \rightarrow 0$. Here we resolve the singularity by introducing a Navier slip boundary condition at the plate, 
\begin{equation}
v_z = l_s \, \frac{\partial v_z}{\partial x}~,
\end{equation}
that is characterized by a slip length $l_s$. Such a slip law has been confirmed experimentally, yielding values for $l_s$ ranging from a single molecular length up to a micron depending on wetting properties of the liquid and roughness of the solid (\cite{BB99}, \cite{CCSC05}, \cite{PHL00}, \cite{TR89}). Different mechanisms releasing the contact line singularity will lead to similar qualitative results, as long as the microscopic and macroscopic lengths remain well separated. 

The lubrication equation with slip boundary condition reads (\cite{ODB97})
\begin{eqnarray}
\partial_t h+ \nabla \cdot \left( h \,{\bf U}\right) &=& 0~, \\
\gamma \nabla \kappa - \rho g {\bf e}_z + \frac{3\eta(U\, {\bf e}_z  - {\bf U})}{h(h+3l_s)} &=& {\bf 0}~.\label{momentumdim}
\end{eqnarray}
Here $U$ is the plate velocity, ${\bf U}(z,y,t)=U_z {\bf e}_z+U_y {\bf e}_y$ is the depth-averaged fluid velocity inside the film, while $\nabla={\bf e}_z\partial_z + {\bf e}_y\partial_y$. The first equation is mass conservation, while the second represents the force balance between surface tension $\gamma$, gravity $\rho g$, and viscosity $\eta$, respectively. We maintain the full curvature expression expression 

\begin{eqnarray}
\label{curvature}
\kappa &=&  \frac{\left(1+ \partial_{y} h^2\right)\partial_{zz}h + \left(1+ \partial_{z} h^2\right)\partial_{yy}h -2\partial_y h \, \partial_z h \, \partial_{yz}h}{(1+\partial_z h^2 + \partial_y h^2)^{3/2}}~,
\end{eqnarray}
which allows a proper matching to the liquid reservoir away from the contact line.  

In the remainder we rescale all lengths by the capillary length $l_\gamma=\sqrt{\gamma/\rho g}$, and all velocities by $\gamma/\eta$, yielding the dimensionless equations
\begin{eqnarray}
\label{continuity}
\partial_t h+ \nabla \cdot \left( h \,{\bf U}\right) &=& 0, \\
\nabla \kappa - {\bf e}_z + \frac{3({\rm Ca}\,{\bf e}_z  - {\bf U})}{h(h+3l_s)} &=& {\bf 0}.
\label{momentum}
\end{eqnarray}
The timescale in this equation thus becomes $\eta l_\gamma /\gamma$.

\subsection{Boundary conditions}

We now have to specify boundary conditions at the liquid reservoir and at the contact line -- see Fig.~\ref{fig.sketch}. Far away from the plate, $h\rightarrow \infty$, the free surface of the bath is unperturbed by the contact line. Defining the vertical position of the bath at $z=0$, we can thus impose the asymptotic boundary conditions as $z\rightarrow 0$
\begin{eqnarray}\label{bcbath}
\partial_z h &=& -\infty, \nonumber \\
\partial_y h &=& 0, \nonumber \\
\kappa &=& 0.
\end{eqnarray}
We impose at the contact line, at $z=z_{\rm cl}$, that
\begin{eqnarray}\label{bccontactline}
h &=& 0, \nonumber  \\
|\nabla h| &=&  \tan \theta_{\rm cl}, \nonumber \\
h {\bf U} &=& {\bf 0}.
\end{eqnarray}
The first condition determines the position of the contact line, while the third condition ensures that no liquid passes the contact line. This condition is not trivial, since the equations admit solutions where the liquid velocity diverges as $\sim 1/h$. The second condition imposes the microscopic contact angle, $\theta_{\rm cl}$, emerging from the force balance at the contact line. This condition is actually hotly debated: for simplicity it is often assumed that this microscopic angle remains fixed at its equilibrium value (\cite{H01,E04}), but measurements have suggested that this angle varies with ${\rm Ca}$ (\cite{RGW04}). We will limit ourselves to presenting an argument in favour of a fixed microscopic angle. 
The boundary condition arises at a molecular scale, $l_{\rm VdW}$, at which the fluid starts to feel the van der Waals forces exerted by the substrate. This effect can be incorporated by a disjoining pressure $A/h^3$, where the Hamaker constant $A \propto \gamma l_{VdW}^2$ (\cite{I92}). At $h=l_{\rm VdW}$, this yields a contribution of the order $A h'/ l_{\rm VdW}^2$ in Eq.~(\ref{momentumdim}). Taking $l_{\rm VdW}\sim l_s$, the viscous stresses will have a relative influence on the disjoining term, and thus on the boundary condition, of the order of ${\rm Ca} \sim 10^{-2}$, so that the contact angle should roughly remain within $1\%$ of its equilibrium value. 
This analysis of the microscopic contact angle will be extended and compared to novel experimental results in a forthcoming paper (\cite{JFM2}).

\subsection{The basic profile and the bifurcation diagram}

We first solve for the basic profile $h_0(z)$, corresponding to a stationary meniscus that is invariant along $y$. From continuity we find that $h U_z$ is constant, which using the boundary condition Eq.~(\ref{bccontactline}) yields $U_z=0$. The momentum balance Eq.~(\ref{momentum}) for $h_0(z)$ thus reduces to 
\begin{equation}\label{basicprofile}
\kappa_0' = 1 - \frac{3{\rm Ca} }{h_0\left( h_0 + 3l_s \right)},
\end{equation}
with 
\begin{equation}
\kappa_0 = \frac{h_0''}{\left( 1 + h_0'^2\right)^{3/2}}.
\end{equation}
Close to the bath, i.e. $z\approx 0$, the height of the interface becomes much larger than the capillary length, so that one can ignore the viscous term. The asymptotic solution of Eq.~(\ref{basicprofile}) near $z=0$ thus simply corresponds to that of a static bath,
\begin{eqnarray}
h_0 &=& -\ln z/c , \nonumber \\
h_0' &=& -\frac{1}{z}, \nonumber \\
\kappa_0 &=& z,
\end{eqnarray}
which indeed respects the boundary conditions of Eq.~(\ref{bcbath}). This solution has one free parameter, $c$, that can be adjusted to fullfill the remaining boundary condition at the contact line, $h_0' = -\tan \theta_{\rm cl}$. It is not entirely trivial that this uniquely fixes the solution since Eq.~(\ref{basicprofile}) degenerates for $h_0=0$. One can show that the asymptotic solution for small $Z=z_{\rm cl} - z$ reads~(\cite{BSB03})

\begin{equation}
h_0 = \tan \theta_{\rm cl} \, Z - 
\frac{(1+ \tan^2 \theta_{\rm cl})^{3/2}{\rm Ca}}{2l_s\tan \theta_{\rm cl}} \, Z^2 \ln Z + \tilde{c} Z^2~,
\end{equation}
where $\tilde{c}$ is the sole degree of freedom. The expansion can be continued to arbitrary order once $\theta_{\rm cl}$ and $\tilde{c}$ are fixed.
\begin{figure}
\includegraphics{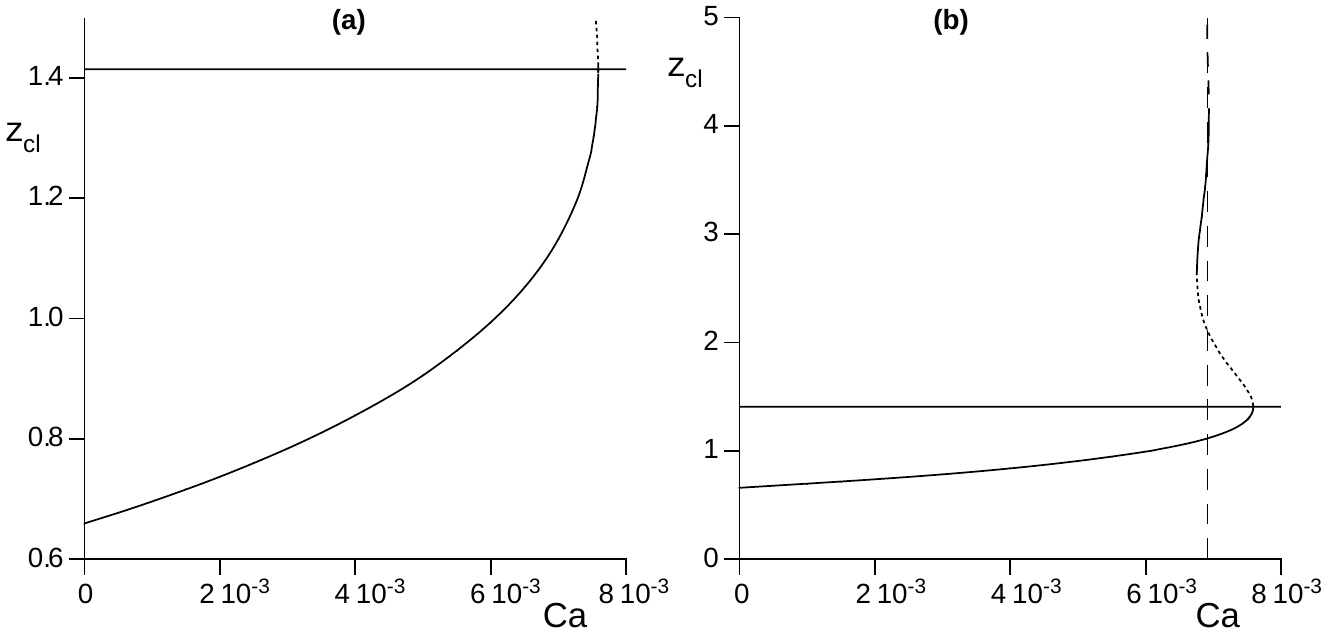} 
\centering 
\caption{(a) Contact line position $z_{\rm cl}$ at equilibrium as a function of the capillary number $Ca$ for fixed parameters $\theta_{\rm cl}=51.5^\circ$ and $l_s=5\cdot 10^{-7}$. Stationary solutions cease to exist above a critical value ${\rm Ca}_c=0.00759\cdots$. The horizontal bar denots $z_{\rm cl}=\sqrt{2}$. (b) Same as (a), but now showing the full range of $z_{\rm cl}$. The solutions undergo a sequence of saddle-node bifurcations with ultimately $z_{\rm cl}\rightarrow \infty$, with a corresponding ${\rm Ca}^* = 0.00693\cdots$ (dashed line).} 
\label{fig.zcl} 
\end{figure} 

So, the basic profile $h_0(z)$ is indeed determined by the microscopic parameters $\theta_{\rm cl}$ and $l_s$, and the (experimental) control parameter ${\rm Ca}$. This is illustrated in Fig.~\ref{fig.zcl}a, showing $z_{\rm cl}({\rm Ca})$ for fixed parameters $\theta_{\rm cl}=51.5^\circ$ and $l_s=5\cdot 10^{-7}$. Similar to \cite{H01} and \cite{E04}, we find that stationary meniscus solutions only exist up to a critical value ${\rm Ca}_c$. Beyond this capillary number the interface has to evolve dynamically and a liquid film will be deposited onto the plate. One can use Fig.~\ref{fig.zcl}a to extract the apparent contact angle $\theta_a$, via Eq.~(\ref{zcl}). The critical capillary number is attained when $z_{\rm cl}=1.4076\cdots$, which is very close to $\sqrt{2}=1.4142\cdots$. This confirms the predictions by~\cite{E04} that stationary solutions cease to exist at a zero apparent contact angle. This slight difference from $\sqrt{2}$ is due to the fact that Eggers's asymptotic theory becomes exact only in the limit where $l_s\rightarrow 0$, so that minor deviations can indeed be expected. Other values of $\theta_{\rm cl}$ and $l_s$ lead to very similar curves, always with a transition at $z_{\rm cl}\simeq\sqrt{2}$, but with shifted values of ${\rm Ca}_c$. This critical value roughly scales as ${\rm Ca}_c \propto \theta_{\rm cl}^3/\ln l_s^{-1}$ (\cite{E04,E05}).

\begin{figure}
\includegraphics{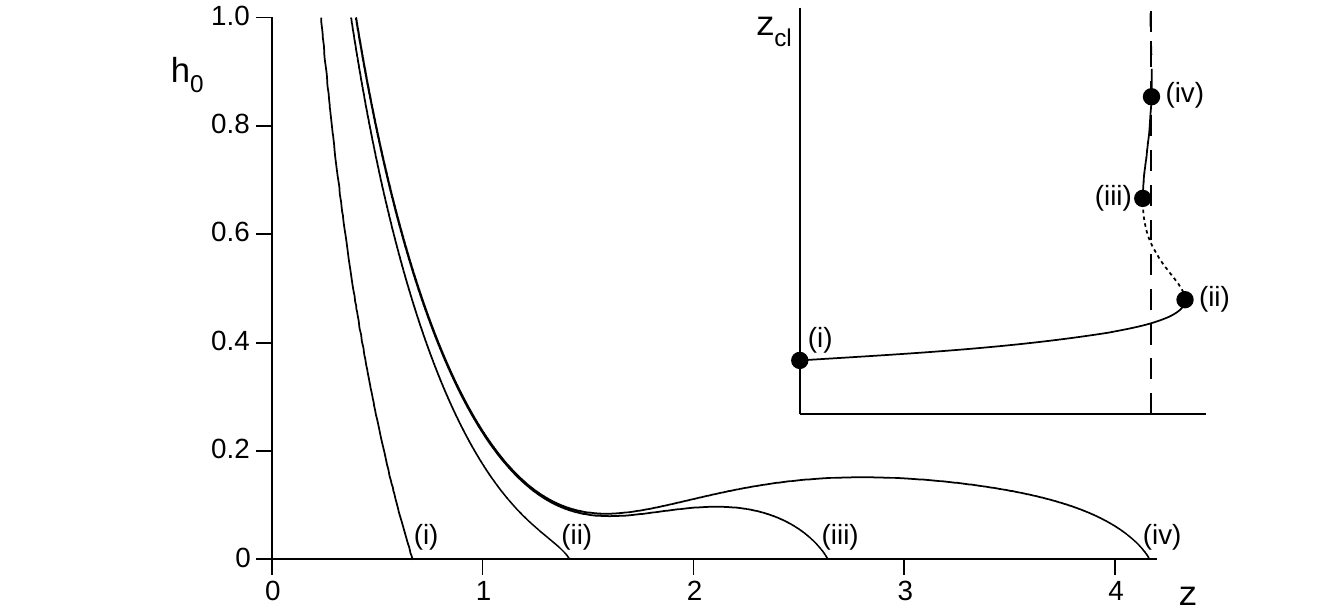} 
\centering 
\caption{Various basic solutions $h_0(z)$ along the bifurcation diagram of Fig.~\ref{fig.zcl}, see inset.} 
\label{fig.solutions} 
\end{figure}

In fact, the existence of a maximum capillary number is due to a saddle-node bifurcation, which originates from the coincidence of a stable and and unstable branch (this will be shown in more detail in Sec.~\ref{sec.results}). As can be seen from Fig.~\ref{fig.zcl}b, there is a branch that continues above $z_{cl}=\sqrt{2}$. Surprisingly, these solutions subsequently undergo a series of saddle-node bifurcations, with capillary numbers oscillating around a new ${\rm Ca}^*$. This asymptotically approaches a solution of an infinitely long flat film behind the contact line. Figure~\ref{fig.solutions} shows the corresponding profiles $h_0$, and illustrates the formation of the film. This film is very different from the so-called Landau-Levich film, which was computed in a classic paper \cite{LL42}). The Landau-Levich solution is much simpler in the sense that it does not involve a contact line and does not display the non-monotonic shape shown in Fig.~\ref{fig.solutions}. The difference markedly shows up in the thickness of the film: while the Landau-Levich film thickness scales as ${\rm Ca}^{2/3}$, the film with a contact line has a thickness $h_\infty = \sqrt{3{\rm Ca}^*}$. Note that very similar film solutions were already identified by \cite{H01} and more recently by \cite{ME05} in the context of Marangoni-driven flows.

These new film solutions have indeed been recently experimentally, as transient states in the deposition of the Landau-Levich film (\cite{SDFA06}). In fact, the transition towards entrainment was observed to coincide at ${\rm Ca}^*$, hence well {\em before} the critical point ${\rm Ca}_c$ and with $\theta_a \neq 0$. For fibres, on the other hand, the condition of vanishing contact angle has been observed experimentally by \cite{S91}. We come back to this issue at the end of the paper. 

\subsection{Physical meaning of the apparent contact angle}

\begin{figure}
\includegraphics{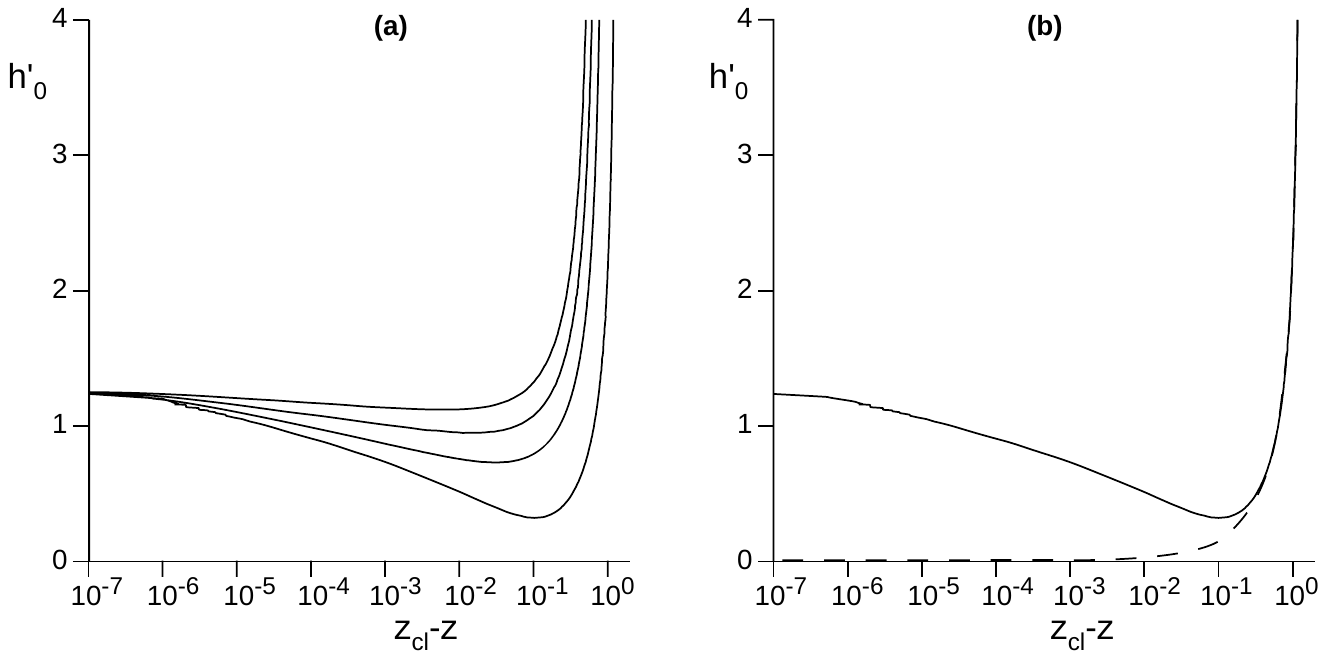} 
\centering 
\caption{(a) Variation of the slope $h_0'$ as a function of the distance to the contact line, for different values of ${\rm Ca}$. (b) The profile $h_0'$ for the critical solution (solid line), compared to the static bath with $\theta_a=0$ (dashed line).} 
\label{fig.hprime} 
\end{figure}

From the definition of Eq.~(\ref{zcl}), it is clear that the apparent contact angle represents an extrapolation of the large scale profile using the static bath solution. In reality, however, the interface profile is strongly curved near the contact line and the contact angle increases to a much larger $\theta_{\rm cl}$. This has been shown in Fig.~\ref{fig.hprime}, revealing the logarithmic evolution of the interface slope close to the contact line. This is different from the static bath solutions, for which the slope decreases monotonically when approaching the contact line (Fig.~\ref{fig.hprime}b, dashed curve). Another way to define a typical contact angle in the dynamic situation could thus be to use the inflection point, which yields the minimum slope of the interface. However, when using $\theta_a({\rm Ca})$ as an asymptotic matching condition for an outer scale solution, like in a quasi-static theory, it is clear that it only makes sense to use the extrapolated version. 

\section{Linear stability within the hydrodynamical model}\label{sec.linstab}

We now turn to the actual linear stability analysis within the hydrodynamic model. This section poses the mathematical problem and addresses some technical issues related to asymptotic boundary conditions. The numerical results will be presented in Sec.~\ref{sec.results}.

\subsection{Linearized equation and boundary conditions}

We linearize Eqs.~(\ref{continuity},\ref{momentum}) about the basic profile $h_0(z)$, writing
\begin{eqnarray}
\label{perturbationfull}
h(z,y,t)&=&h_0(z) + \epsilon\, h_1(z)\,e^{-\sigma t + iqy}, \\ 
\kappa(z,y,t) &=& \kappa_0(z) + \epsilon \, \kappa_1(z) \,e^{-\sigma t + iqy}, \\
U_z(z,y,t) &=& \epsilon \, U_{z1}(z) \,e^{-\sigma t + iqy}, \\
U_y(z,y,t) &=& \epsilon \, U_{y1}(z) \,e^{-\sigma t + iqy}.
\end{eqnarray}
Here we used that the basic velocity ${\bf U}_0={\bf 0}$, so that the velocity is of order $\epsilon$ only. The linearized equation is not homogeneous in $z$, due to $h_0(z)$, so the eigenmodes are nontrivial in the $z$ direction. From the $y$-component of Eq.~(\ref{momentum}), one can eliminate $U_{y1}$ in terms of $\kappa_1$, as 
\begin{equation}
U_{y1}(z) = \frac{1}{3}\, iq h_0\left( h_0 + 3l_s \right) \,\kappa_1(z).
\end{equation}
It is convenient to introduce the variable 
\begin{equation}
F_1(z) = h_0(z) \, U_{z1}(z), 
\end{equation}
which represents the flux in the $z$ direction at order $\epsilon$ (the zeroth order flux being zero). Writing the vector 
\begin{equation}
\vec{X} =  \left(\begin{array}{c} h_1  \\ h_1'  \\ \kappa_1 \\ F_1 \end{array} \right) ,
\end{equation}
one can cast the linearized equation for the eigenmode as
\begin{equation}
\label{linearized}
d_z \vec{X} = {\cal A}\vec{X},
\end{equation}
where $d_z$ denotes the derivative with respect to $z$ and the right hand side is a simple matrix product. From linearization of Eqs.~(\ref{curvature}),~(\ref{continuity}),~and~(\ref{momentum}) one finds
\begin{eqnarray}
{\cal A} =\left(\begin{array}{cccc} 0 & 1 & 0 & 0 \\ 
&&&\\
q^2(1+h_0'^2) & 3h_0' \kappa_0(1+h_0'^2)^{1/2} & (1+h_0'^2)^{3/2} & 0 \\
&&&\\
\frac{3 {\rm Ca}}{h_0^2\left(h_0 + 3l_s \right)}(2-\frac{3l_s}{h_0 + 3l_s}) & 0 & 0 & \frac{3}{h_0^2\left(h_0 + 3l_s \right)} \\
&&&\\
\sigma & 0 &  q^2 h_0^2(h_0 + 3l_s)/3   & 0 \end{array}\right). \nonumber 
\end{eqnarray}
The eigenmodes and corresponding eigenvalues $\sigma_q$ of this linear system are determined through the boundary conditions. At the contact line, we have to obey the boundary conditions of a microscopic contact angle $\tan \theta_{\rm cl}$ and a zero flux 
(see Eq.~(\ref{bccontactline})). To translate this in terms of the linearized variables, we have to evaluate $|\nabla h|$ at the position of the contact line $z_{\rm cl}+\Delta z$. Along the lines of App.~\ref{app.quasistatic} one finds $\Delta z = \epsilon e^{-\sigma t +iqy} h_1/\tan \theta_{\rm cl}$. Linearizing $|\nabla h|$ then yields the boundary conditions for the eigenmode
\begin{eqnarray}
\label{bcmode1}
h_1' &=& - \frac{\kappa_0 \left( 1 + h_0'^2\right)^{3/2}}{\tan \theta_{\rm cl} } h_1, \\
F_1&=&0 \label{bcmode2}.
\end{eqnarray}
At the side of the bath, $z\rightarrow 0$, the conditions become
\begin{eqnarray}\label{bcbath2}
iq h_1 &=& 0 \quad \Rightarrow \quad q=0 \quad \vee \quad h_1=0, \\
\kappa_1 &=& 0 .\label{bcbath3}
\end{eqnarray}
Below we identify the two relevant asymptotic behaviors at the bath respecting these boundary conditions.

\subsection{Shooting: asymptotic behaviors at bath}

The strategy of the numerical algorithm is to perform a shooting procedure from the bath to the contact line, where we have to obey the two conditions Eq.~(\ref{bcmode1}) and~(\ref{bcmode2}). We thus require two degrees of freedom, one of which 
is the sought for eigenvalue $\sigma$. Since the problem has been linearized, the amplitude of a single asymptotic solution does not represent a degree of freedom: one can use the relative amplitudes of two asymptotic solutions as the additional parameter to shoot towards the contact line. We thus need to identify two linearly independent solutions that satisfy the boundary boundary conditions (\ref{bcbath2},\ref{bcbath3}). 

There are two asymptotic solutions of the type:
\begin{eqnarray}
h_1 &=& z^\alpha \left( 1 + \frac{3{\rm Ca}}{2(\alpha+1)^2} \,\frac{1}{\ln^2 (z/c)} \right) \nonumber \\
h_1' &=& \alpha z^{\alpha-1} \left( 1 + \frac{3{\rm Ca}}{2(\alpha+1)^2 } \,\frac{1}{\ln^2 (z/c)} \right)\nonumber \\
\kappa_1 &=& \frac{-6 {\rm Ca}}{\alpha+1} \,\frac{z^{\alpha+1}}{\ln^3 (z/c)} \nonumber \\
F_1 &=& \frac{\sigma}{\alpha+1}\, z^{\alpha+1}~.
\end{eqnarray}
These exist for the two roots of 
\begin{equation}
\alpha^2 + 2\alpha - q^2 =0 \quad \Rightarrow \quad \alpha_\pm = \pm \sqrt{1+q^2} - 1~.
\end{equation}
Since we want $h_1$ to be bounded, only the solution $\alpha_+$ is physically acceptable. Interestingly, the mode $h_1\propto z^{\alpha_+}$ precisely has the well-known Laplacian structure of $\exp(-qx+iqy)$ when transformed in the frame where the bath is horizontal -- see Appendix~\ref{app.tetetourne}. This mode thus corresponds to a zero curvature perturbation of a static bath, with no liquid flow. Indeed, no flux crosses the bath since $F_1\rightarrow 0$. 

However, the motion of the contact line implies that liquid is being exchanged with the liquid reservoir, so we require an asymptotic solution that has a nonzero value of $F_1$. We found that the corresponding mode has the following structure: 
\begin{eqnarray}
h_1 &=&  -\frac{q^2 {\cal L}(z)}{z} + \frac{1}{\ln^3 (z/c)} \nonumber \\
h_1' &=& \frac{q^2 {\cal L}(z)}{z^2} + \frac{q^2}{z\ln^3 (z/c)} - \frac{3}{z \ln^4 (z/c)} \nonumber \\
\kappa_1 &=& q^2 (1+q^2) {\cal L}(z)
\nonumber \\
F_1 &=& \frac{1}{3} q^2(1+q^2)~.
\end{eqnarray}
where ${\cal L}$ stands for
\begin{equation}
{\cal L}(z)= \int_0^z dt\, \frac{-1}{\ln^3 (t/c)}~. 
\end{equation}
This integral can be rewritten in terms of a logarithmic integral using partial integration, but this does not yield a simpler expression. 

For completeness, let us also provide the fourth asymptotic solution of this fourth order system:
\begin{eqnarray}
h_1 &=& \frac{1}{z} \left( 1 - \frac{3{\rm Ca}}{1+q^2} \,\frac{1}{\ln^2 (z/c)} \right) \nonumber \\
h_1' &=& -\frac{1}{z^2}\left( 1 - \frac{3{\rm Ca}}{1+q^2} \,\frac{1}{\ln^2 (z/c)} \right)\nonumber \\
\kappa_1 &=& -(1+q^2)\left(  1 - \frac{3{\rm Ca}}{1+q^2} \,\frac{1}{\ln^2 (z/c)}    \right) \nonumber \\
F_1 &=& \sigma \ln (z/c)~,
\end{eqnarray}
which clearly violates the boundary conditions Eq.~(\ref{bcbath2},\ref{bcbath3}).

To summarize, there are two asymptotic solutions that are compatible with the boundary conditions at the bath. Their relative amplitudes can be adjusted to satisfy one of the two boundary conditions at the contact line. The numerical shooting procedure allows finding the eigenvalue $\sigma_q$ for which also the second boundary condition is obeyed. 

\section{The dispersion relation}\label{sec.results}

\subsection{Numerical results}

Let us now discuss the dispersion relation of contact line perturbations obtained within the hydrodynamic framework. For fixed microscopic parameters, the relaxation rate depends on the capillary number ${\rm Ca}$ and the dimensionless wave number $q$ that has been normalized by the capillary length. This relation will be represented by the function $\sigma_q({\rm Ca})$, which has the dimension of the inverse time-scale $\gamma/(\eta l_\gamma)$. From the definition Eq.~(\ref{perturbationfull}), it follows that $\sigma$ is positive for stable solutions. 

\begin{figure}
\includegraphics{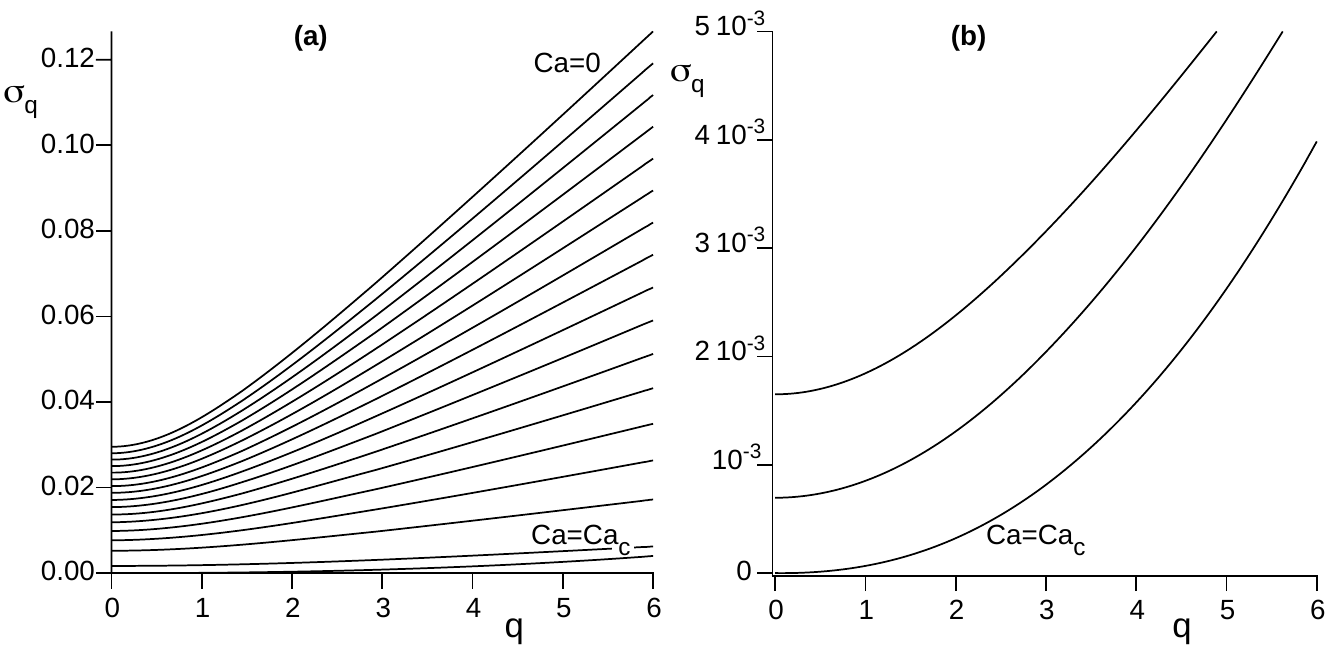} 
\centering 
\caption{Dispersion relation obtained numerically. (a) Relaxation rates $\sigma$ as a function of $q$, in units of $\gamma/(\eta l_\gamma)$ and $1/\l_\gamma$ respectively. The various curves correspond to values of ${\rm Ca}$ ranging from $0$ to $7.5~10^{-3}$ increasing by steps of $0.5~10^{-3}$, plus ${\rm Ca}_c$. (b) Same but close to the critical capillary number: from top to bottom, ${\rm Ca}=0.0075$, ${\rm Ca}=0.0075687$ and ${\rm Ca} \simeq {\rm Ca}_c=0.00758751$.} 
\label{fig.sigmaq} 
\end{figure}

The dispersion relations are summarized by Fig.~\ref{fig.sigmaq}, displaying $\sigma_q$ for various values of ${\rm Ca}$. For values well below the critical speed ${\rm Ca}_c$, one finds that the relaxation increases with $q$, in a manner consistent with the quasi-static prediction that $\sigma \propto |q|$ for large $q$. The crossover towards this linear scaling happens around $q \approx 1$, and is thus governed by the capillary length. 

Close to the critical point, however, we find two unexpected features. First, it is clear from Fig.~\ref{fig.sigmaq}b that the linear regime disappears, or lies outside the range of our curves. We have not been able to extend the numerical calculation to larger values of $q$ due to intrinsic instability of the numerical algorithm (the presented curves have arbitrary precision). Hence, the crossover value for $q$, denoted by the inverse wavelength $1/\lambda_{\rm cut}$, increases dramatically close to the transition. Second, we observe a vanishing relaxation rate for the mode $q=0$ at ${\rm Ca}_c$, or equivalently a diverging relaxation time. However, the rates at {\em finite} wavelengths remain nonzero at the transition. This is in contradiction with the quasi-static theory, Eq.~(\ref{ratiostatic}), suggesting that $\sigma_q$ vanishes at all length scales at the transition. 

To characterize these behaviors in more detail, it is convenient to use an empirical relation for the numerical curves (\cite{O92}), 

\begin{equation}\label{fit}
\sigma_q \simeq \sigma_0 + \sigma_\infty \, \left( 
\frac{\sqrt{1+\left(q \lambda_{\rm cut} \right)^2}-1}{\lambda_{\rm cut}} 
\right)~.
\end{equation}
This form contains the two main features of the dispersion: the relaxation rate for the zero mode $\sigma_0({\rm Ca})$, and the prefactor in the linear regime $\sigma_\infty({\rm Ca}) \equiv \lim_{q\rightarrow \infty} \sigma_q/q$ already defined in Eq.~(\ref{sigmainf}). The cut-off length $\lambda_{\rm cut}$ then characterizes the cross-over between the two regimes. The quasi-static prediction would be that $\lambda_{\rm cut} \approx 1$ and $\sigma_\infty \propto \sigma_0$, see Eq.~(\ref{ratiosigmainf}). We have found that Eq.~(\ref{fit}) provides an excellent fit for all data. Only close to the critical point, where the linear regime is no longer observed within our numerical range, the values of $\lambda_{\rm cut}$ and $\sigma_\infty$ are slightly dependent on the choice for the fit. The result for $\sigma_0$ is completely independent of this choice.

\begin{figure}
\includegraphics{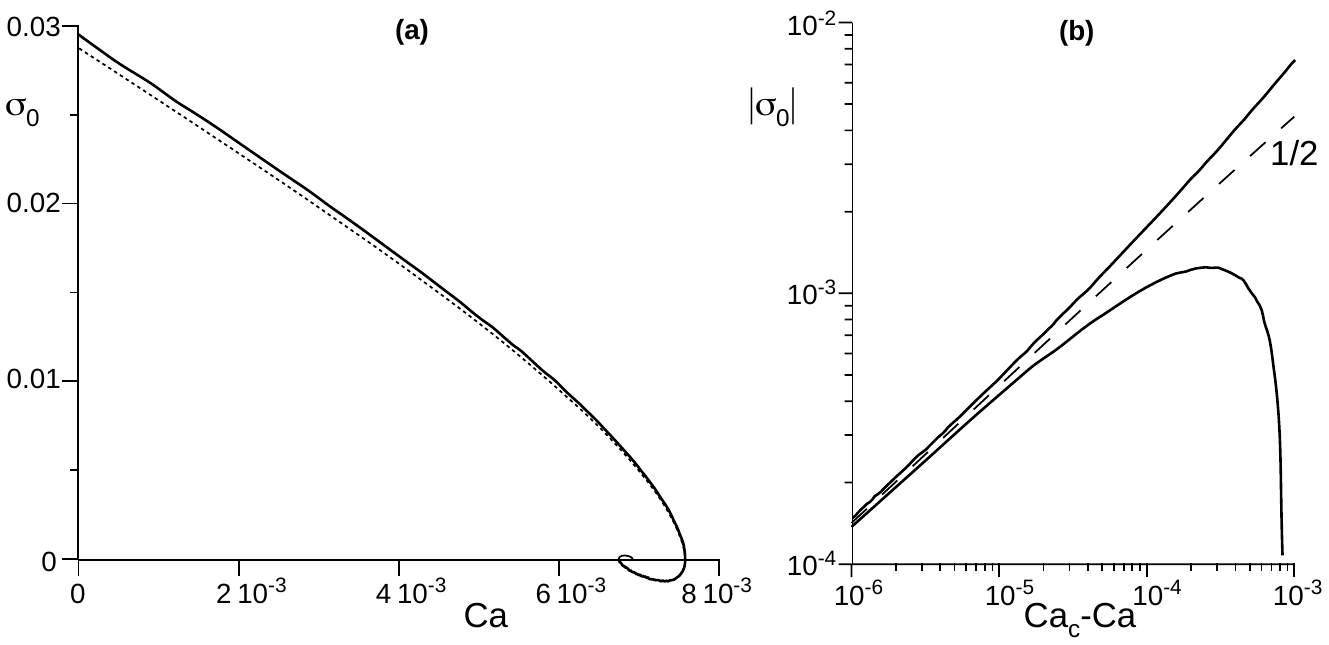} 
\centering 
\caption{(a) Zero mode relaxation rate $\sigma_0$ as a function of  ${\rm Ca}$ (solid line). The dashed line shows the quasi-static approximation. (b) Same but plotted in Log-Log coordinates as a function of  ${\rm Ca_c-Ca}$.} \label{fig.sigmaq0ca} 
\end{figure}

Let us first follow the relaxation of the mode $q=0$ as a function of ${\rm Ca}$. Figure~\ref{fig.sigmaq0ca}a shows that $\sigma_0$ decreases with ${\rm Ca}$, so that the relaxation is effectively slowed down. When approaching the critical point, this stable branch actually merges with the first unstable branch shown in Fig.~\ref{fig.zcl}b, the latter giving negative values for $\sigma_0$. As a consequence, the relaxation rate has to change sign at ${\rm Ca}_c$, so that $\sigma_0=0$ at this point. The graph in Fig.~\ref{fig.sigmaq0ca}b shows that this the relaxation time diverges as $\sigma_0^{-1}\propto 1/\sqrt{{\rm Ca}_c-{\rm Ca}}$. As we argue below, this behavior is a fingerprint of a saddle-node bifurcation. This scenario is repeated when following the higher branches of Fig.~\ref{fig.zcl}b. Indeed, one finds a succession of saddle-node bifurcations at which $\sigma_0$ changes sign. In Fig.~\ref{fig.sigmaq0ca}a this manifests itself as an inward spiral, so that the solution with $z_{\rm cl}\rightarrow \infty$ has $\sigma=0$. The dotted curve in Fig.~\ref{fig.sigmaq0ca}a has been obtained from the quasi-static prediction, Eq.~(\ref{q0fromzcl}), relating $\sigma_0$ to the curve $z_{\rm cl}({\rm Ca})$ displayed in Fig.~\ref{fig.zcl}b: the agreement is excellent.

\begin{figure}
\includegraphics{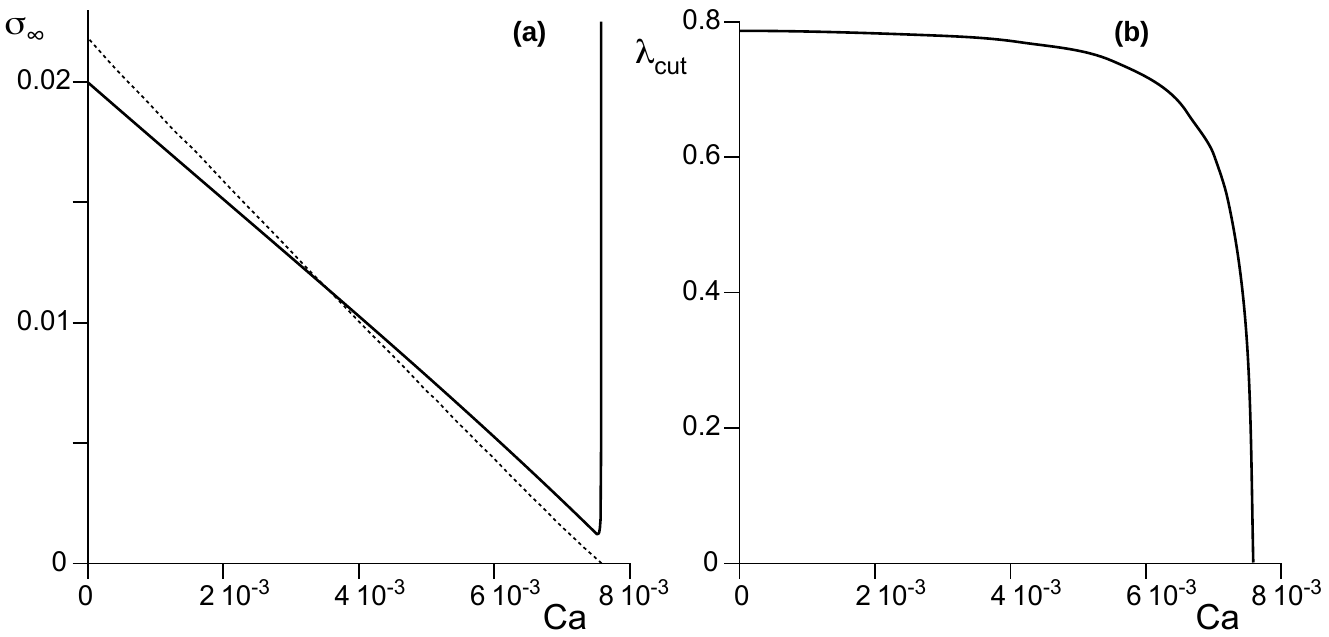} 
\centering 
\caption{(a) Asymptotic relaxation rate $\sigma_\infty \equiv \lim_{q\rightarrow \infty} \sigma_q/q$ as a function of the capillary number ${\rm Ca}$ (solid line). The dotted line shows the quasi-steady prediction, Eq.~(\ref{qfinitestatic}). (b) Crossover wavelength $\lambda_{\rm cut}$ as a function of capillary number ${\rm Ca}$.} 
\label{fig.sigmaqca} 
\end{figure}

This agreement is in striking contrast to the discrepancy at small wavelengths. These are represented in Fig.~\ref{fig.sigmaqca}a through $\sigma_\infty({\rm Ca})$. The comparison with quasi-static theory (dotted line), reveals a significant quantitative disagreement for all ${\rm Ca}$. However, the most striking feature is that $\sigma_\infty$ diverges near the transition. This suggests that for large $q$ the relaxation rates increase faster than linearly, so that the quasi-static theory breaks down even qualitatively. A direct consequence is then that $\lambda_{\rm cut}\rightarrow 0$, as can be seen from Fig.~\ref{fig.sigmaqca}b. At the critical point Eq.~(\ref{fit}) reduces to $\sigma_q \simeq \sigma_\infty \lambda_{\rm cut} \, q^2/2$ for small $q$, so that $\lambda_{\rm cut}\propto 1/\sigma_\infty$. These results underline the qualitative change when approaching ${\rm Ca}_c$.

\subsection{Interpretation}

We propose the following interpretation for the behavior near the critical point. We have seen that the $q=0$ mode is well described through a standard saddle-node bifurcation, which has the normal form
\begin{equation}\label{normal1}
\frac{dA}{dt} = \mu - A^2~.
\end{equation}
For positive $\mu$, this equation has two stationary solutions, namely $A_\pm=\pm \sqrt{\mu}$. Linear stability analysis around these solutions shows that the $A_+$ solutions are stable while the $A_-$ are unstable, and the corresponding relaxation rates scale as $\sigma=\pm 2 \sqrt{\mu}$. So indeed, our numerical results for $q=0$ are  described by the saddle-node normal form, when taking $\mu\propto {\rm Ca} - {\rm Ca}_c$ and $A=\sqrt{2}-z_{\rm cl}$.

Making an expansion around the critical point that incorporates slow spatial variations in $y$, one would expect the following structure
\begin{equation}
\label{normal2}
\frac{\partial A}{\partial t} = \mu - A^2 + D \frac{\partial^2 A}{\partial y^2} ~.
\end{equation}
Due to the symmetry $y \rightarrow -y$, the single derivative of $y$ can never emerge. This then yields a dispersion relation
\begin{equation}
\sigma = \sigma_0 + D q^2 + {\cal O}(q^4)~.
\end{equation}
This explains the observation that for finite $q$ the relaxation rates remain finite at the critical point, even though $\sigma_0=0$. When comparing to Eq.~(\ref{fit}), one finds that $D=\sigma_\infty \lambda_{\rm cut}/2$. This value decreases with ${\rm Ca}$ but remains finite at the transition. Interestingly, however, the dependence $D({\rm Ca})$ appears to extrapolate to zero only about $1\%$ beyond ${\rm Ca}_c$.

\section{Discussion}\label{sec.discussion}

We have performed a hydrodynamic calculation of perturbed receding contact lines, in which viscous dissipation has been taken into account on all length scales (from molecular to macroscopic). This goes beyond earlier work by Golestanian \& Raphael and \cite{NV03}, in which all dissipation was assumed to be localized at the contact line and described by an apparent (macroscopic) contact angle $\theta_a$. 

In the first part of the paper we have revealed the bifurcation diagram for straight contact lines, which turns out to be much richer than expected from the simplified quasi-static approach. Instead of a single saddle-node bifurcation at the critical capillary number ${\rm Ca}_c$, we find a discrete series of such bifurcation points converging to a second threshold capillary number ${\rm Ca}^*$ (Fig.~\ref{fig.zcl}). Interestingly, the latter solutions have been observed experimentally as transient states towards liquid deposition (\cite{SDFA06}). These experiments showed that the wetting transition occurs already at ${\rm Ca}^*$, and hence before the critical value ${\rm Ca}_c$ at which stationary menisci cease to exist. Since we have found the lower branch of Fig.~\ref{fig.zcl} to be linearly stable at all length scales, this subcritical transition has to be mediated by some (unknown) nonlinear mechanism. Let us note that similar experiments using thin fibres instead of a plate suggest that it actually is possible to approach the critical point (\cite{S91}). It would be interesting to investigate the bifurcation diagram as a function of the fibre radius $r$, where the present work represents the limit $r/l_\gamma \rightarrow \infty$.

The second part concerned the relaxation of perturbed contact lines. At long wavelenghts, $q l_\gamma \ll 1$, we have found that the relaxation obtained in the hydrodynamic calculation is very close to the quasi-static prediction. The quasi-static model is based upon the equilibrium contact line position at steady-state, as a function of the capillary number: it treats the perturbations as a small displacement of the contact line, $\Delta z_{\rm cl}$, that induces a change in the contact line velocity $\sim d {\rm Ca}/d z_{\rm cl}$. A positive (negative) derivative indicates that the contact line is stable (unstable). This argument does not involve the apparent contact angle: it holds as long as the interface profile relaxes adiabatically along {\em stationary} or {\em steady} meniscus solutions. The long wavelength theory therefore relies on a "quasi-steady" assumption, and not so much on the interface being nearly at equilibruim (quasi-static). We wish to note that the physics is slightly different for a contact line on a horizontal plane, for which there is no equilibrium position due to gravity. For an infinite volume translational invariance implies that $\sigma_0=0$ (\cite{SOK87}), while drops of finite volume has a finite resistance to long wave-length perturbations. This nicely illustrates the importance of the outer geometry. 

For small wavelengths, $q l_\gamma \gg 1$, we found that the quasi-static theory breaks down. Away from the critical point we still observe the scaling $\sigma \propto |q|$, as proposed by \cite{JdG84}. This scaling reflects the "elasticity" of contact lines, representing an increase of surface area, and thus of the surface free energy, proportional to $|q|$. Quantitatively, however, the quasi-static approximation is not able to capture the hydrodynamic calculation. The disagreement becomes even {\em qualitative} close to the critical point: finite wavelength perturbations do not develop the diverging relaxation times predicted by \cite{GR01a}. Also, the scaling $\sigma \propto |q|$ is found to cross over to a quadratic scaling $\sigma \propto q^2$. 

These results have a clear message: viscous effects have to be treated explicitly when describing spatial structures below the capillary length. Namely, the viscous term in Eq.~(\ref{momentumdim}) becomes at least comparable to gravity at this scale. Therefore one can no longer assume that viscous effects are localized in a narrow zone near the contact line: the perturbations become comparable to the size of this viscous regime. However, even if contact line variations are slow, a complete description still requires a prediction for $z_{\rm cl}({\rm Ca})$, or equivalently $\theta_a({\rm Ca})$. As was shown by \cite{E04}, this relation is not geometry-independent so one can never escape the hydrodynamic calculation. 

Our findings provide a detailed experimental test that, on a quantitative level, are relatively sensitive to the microscopic physics near at the contact line. In our model we have used a simple slip law to release the singularity, but a variety of other mechanisms have been proposed previously. The other model parameter is the microscopic contact angle $\theta_{\rm cl}$, which we have simply taken constant in our calculations. In a forthcoming paper we present experimental results and show to what extent the model is quantitatively accurate for the dynamics of contact lines. 

{\em Acknowledgements --} 
We wist to thank J. Eggers for fruitful discussions and P. Brunet for useful suggestions on the manuscript. JHS acknowledges financial support by a Marie Curie European Fellowship FP6 (MEIF-CT2003-502006).

\appendix 
 
\section{Perturbation of the static bath away from the plate}
\label{app.tetetourne}

At large distances from the plate, the behavior of the static bath is more conveniently described through the function 
$z_{\rm surface}(x,y)=\hat{z}(x,y)$. Denoting $x$ positive away from the plate, we find asymptotically that $\hat{z}=\hat{z}'=\hat{z}''=0$ as $x \rightarrow \infty$. The equation for the static interface then simplifies to 
\begin{equation}
\nabla^2 \hat{z} = \hat{z},
\end{equation}
where we have put $l_\gamma=1$. The basic profile is simply exponential
\begin{equation}
\hat{z}_0=Ae^{-x},
\end{equation}
while transverse perturbations $e^{iqy}$ decay along $x$ as
\begin{equation}
\hat{z}_1 = e^{-\sqrt{1+q^2} x}.
\end{equation} 
We can thus write 
\begin{eqnarray}\label{bla}
\hat{z} &=& Ae^{-x} + \epsilon e^{iqy} (e^{-x})^{{\sqrt{1+q^2}}},
\end{eqnarray}
and compare this to the representation $x_{\rm surface}=h(z,y)$
\begin{equation}
h = - \ln (z/c) + \epsilon e^{iqy} \, h_1(z).
\end{equation}
Inserting this $x=h(z,y)$ in Eq.~(\ref{bla}), and identifying $\hat{z}=z$, we obtain to lowest order in $\epsilon$
\begin{eqnarray}
z &=& \frac{Az}{c}(1-\epsilon e^{iqy}\, h_1(z)) + \epsilon e^{iqy}\left( \frac{Az}{c}\right)^{\sqrt{1+q^2}},
\end{eqnarray}
so that $A/c=1$ and $h_1(z)=z^\alpha$, with $\alpha=\sqrt{1+q^2}-1$. So, the exponential relaxation in the frame $(x,y)$ 
translates into a power law for $h_1(z)$.

\section{Quasi-static approximation}\label{app.quasistatic}

In this appendix we derive the quasi-static results summarized in Sec.~\ref{sec.quasistatic}. To perform a linear stability analysis, we write the interface profile as 
\begin{eqnarray}
\label{perturbation}
h(z,y,t)&=&h_0(z) + \epsilon\, h_1(z)\,e^{-\sigma t + iqy}, \\ 
\kappa(z,y,t) &=& \kappa_0(z) + \epsilon \, \kappa_1(z) \,e^{-\sigma t + iqy},
\end{eqnarray}
where $\kappa$ is twice the mean curvature of the interface. In the quasi-static approach, the basic profile $h_0(z)$ can be solved from a balance between capillary forces and gravity, so that the scale for interface curvatures is the capillary length $l_\gamma=\sqrt{\gamma/\rho g}$. If we consider modulations of the contact line with short wavelengths, $1/q \ll l_\gamma$, one can thus locally treat the unperturbed profile as a straight wedge, $h_0(z)= (z_{\rm cl} - z) \tan \theta_a$, where the position of the contact line is denoted by $z_{\rm cl}$. Since gravity plays no role at these small length scales, one can easily show that the perturbation should have zero curvature, i.e. $\kappa_1(z)=0$. In the limit of small contact angles, for which we can simply write 
\begin{equation}
\kappa_1 \simeq \nabla^2 (h_1(z)\, e^{-\sigma t + iqy})~, 
\end{equation}
one directly finds that $\kappa_1=0$ leads to a perturbation decaying exponentially along $z$, as $h_1(z)=e^{-|q|(z_{\rm cl}-z)}$. The length scale of the perturbation is then simply $1/q$. This can be generalised using the full curvature expression Eq.(~\ref{curvature}): inserting the linearization Eq.~(\ref{perturbation}), and taking 
$\partial_z h_0 =\tan\theta_a$, $\partial_{zz}h_0=0$, one finds
\begin{equation}
\kappa_1 = (\cos \theta_a)^{3/2} \left(\partial_{zz} h_1 - \frac{q^2}{\cos^2 \theta_a} h_1\right).
\end{equation}
Hence, the condition that $\kappa_1=0$ yields:
\begin{equation}
h_1(z)=e^{-|q|\frac{z_{\rm cl}-z}{\cos \theta_a}}~.
\end{equation}

From Fig.~\ref{fig.sketch}a it is clear that the "advanced" part of the contact line, with positive $\Delta z$, has a smaller apparent contact angle than the wedge. The remaining task is to relate the quantities $\epsilon$, $\Delta z$ and $\Delta \theta$, and to impose the correct boundary condition through $\theta_a({\rm Ca})$. We now introduce the representation 
\begin{eqnarray}
h(z,y,t) &=& h_0(z-\Delta z ) + 
\epsilon \hat{h}_1(z-\Delta z) \, e^{-\sigma t + iqy}~, 
\end{eqnarray}
in which the position of the contact line is explicitly shifted to $z=z_{\rm cl}+\Delta z$, so that $\hat{h}_1(z_{\rm cl})=0$. Linearizing this equation around $z=z_{\rm cl}$, this can be written as
\begin{eqnarray}
h(z,y,t) &=& h_0(z) + \epsilon \left( \hat{h}_1(z)e^{-\sigma t+iqy} - \left[\frac{dh_0}{dz}\right]_{z=z_{\rm cl}} \frac{\Delta z}{\epsilon}\right) + {\cal O}(\epsilon^2) \nonumber \\
&=& h_0(z) + \epsilon \left( \hat{h}_1(z) e^{-\sigma t+iqy} + \tan \theta_a \frac{\Delta z}{\epsilon}\right) + {\cal O}(\epsilon^2).
\end{eqnarray}
Comparing to Eq.~(\ref{perturbation}) with $h_1(z_{\rm cl})=1$, one thus finds that to lowest order $\Delta z = \epsilon e^{-\sigma t+iqy} /\tan \theta_a$. Writing $\partial h/\partial z=-(\tan \theta_a + \Delta \tan \theta)$, one furthermore finds from Eq.~(\ref{perturbation}) 
\begin{equation}
\Delta \tan \theta = -\frac{|q|}{\cos \theta_a} \epsilon e^{-\sigma t+iqy} = -|q| \frac{\tan \theta_a}{\cos \theta_a} \Delta z.
\end{equation}
The final step is to use the empirical relation between $\theta_a$ and ${\rm Ca}$ 
to relate the variation in contact angle to a variation in the contact line velocity, $U_{\rm cl}=U-d\Delta z/dt$: 
\begin{eqnarray}
\frac{d\Delta z}{dt}&=& -\frac{\gamma}{\eta}\, \Delta {\rm Ca} = -\frac{\gamma}{\eta}\, \left( \frac{d \tan \theta_a}{d{\rm Ca}} \right)^{-1} \Delta \tan \theta 
\nonumber \\
&=& \frac{|q|\gamma}{\eta}\,\frac{\tan \theta_a}{\cos \theta_a} \, \left( \frac{d \tan \theta_a}{d{\rm Ca}} \right)^{-1} \Delta z.
\end{eqnarray}
This indeed results into an exponential relaxation 
$\Delta z \propto e^{-\sigma t}$ with a relaxation rate given by Eq.~(\ref{qfinitestatic}). 

For contact angles close to $\pi/2$, one can analytically solve the crossover from small to large wavelengths (\cite{NV03}). In this case the basic profile is nearly flat in the $x$ direction. Characterizing the free surface by $z_{\rm surface}(x,y)$, one easily finds that perturbations of the contact line decay along $x$ over a distance $1/\sqrt{q^2+1/l_\gamma^2}$ -- see Appendix~\ref{app.tetetourne}. Hence, 
\begin{equation}\label{vadim}
\frac{\sigma_q}{\sigma_0} \simeq \sqrt{1+(ql_\gamma)^2} \quad {\rm for} \quad  \theta_a \approx \pi/2.
\end{equation}
This is consistent with Eq.~(\ref{ratiostatic}), since $g(\pi/2)= 1$.

\end{document}